# A Comparison of Audio Preprocessing Techniques and Deep Learning Algorithms for Raga Recognition

**Devayani Hebbar**

MIT World Peace University

1032181272@mitwpu.edu.in

**Vandana Jagtap**

MIT World Peace University

vandana.jagtap@mitwpu.edu.in

## 1. ABSTRACT

**Ragas form the foundation for Indian Classical Music. The task of Raga Recognition has gained traction in the Music Information Retrieval community in the recent past, which can be attributed to the nuances of Indian Classical Music that have resulted in a plethora of research problems in Computing. In this work, we used two different digital audio signal processing (DASP) techniques to preprocess audio samples of Carnatic classical ragas that were then processed by various Deep Learning models. Their results were compared in order to infer which DASP technique is better suited to the task of raga recognition. We obtained state of the art results, with our best model reaching a testing accuracy of 98.1%. We also compared each model's ability to distinguish between similar ragas.**

Keywords- *Raga* ● *Swara* ● Spectrogram ● Feature Extraction ● CNN ● ANN ● LSTM

## 2. INTRODUCTION

Much like Western music, the entirety of Indian classical music is composed using a total of 12 musical notes. In Carnatic (South Indian) classical music, the 12 notes are called *swaras*. A raga is a combination of a fixed set of *swaras* arranged in a unique sequence [15]. There are 72 "parent" ragas in Carnatic music. These 72 ragas are heptatonic scales (all containing 7 *swaras* each) and are called *melakartha ragas*. Any given raga has been derived from one of the 72 parent ragas and must contain only subsets of *swaras* present in the parent raga that they come from. [1]. Any composition in Carnatic music must be composed in a specific raga and thus, a raga is essentially a compositional framework. A characteristic feature of Carnatic music is the oscillation between *swaras*- an ornament called the *gamaka*. This sometimes makes the task of identifying each individual swara in a phrase challenging. Further, even though no two ragas can be identical, some ragas have the same set of *swaras* and yet, are considered as two separate ragas due to differences in order of occurrence of the *swaras* [4]. This is what makes Raga Recognition such an important and intriguing task in Music Information Retrieval (MIR).

The objective of this work is not only to examine how different Deep Learning (DL) models perform this task, but also to compare two methods of audio preprocessing: (1) Extracting (numerical) features from raw audio files and using these numerical features as data; (2) Converting the same audio files to image data. The results yielded by the two methods are then compared. Another objective is to observe how well each model is able to distinguish between two different ragas that have the same *swaras* and whether approach (1) or approach (2) is better suited to solving this problem.

## 3. RELATED WORK

Researchers have approached the task of raga recognition in different ways, in terms of both, the method of audio preprocessing as well as the type of ML/DL model. Shah et al. converted their audio data into spectrograms. Prior to this step, they also separated the vocals from the instruments in all the audio files. They then conducted a series of experiments in which the spectrograms were used to train a CNN model whose last fully connected layers were dropped in order to obtain a sequence of feature vectors. This sequence of vectors was then used to train an LSTM model [4]. In [5], the audio data was converted to a different type of image representation. They visualized the variation of the predominant tonic pitch values of each song over time. These pitch vs time graphs were then used as image data to train a 2D CNN model. They also divided their ragas into groups in order to observe their model's performance on ragas with similar or identical *swaras* vs ragas with distinct differences in *swaras*. Their model performed well in the latter case, but could not give satisfactory results in the former. In [2] and [3], a different method for preprocessing audio was used. Feature extraction was performed on the audio data in order to obtain a table of numerical features that would then be used to train ML/DL models. In [2], features of three categories were extracted for a set of *melakartha ragas*: spectral, timbre and tonal. They were then used to train ANN models and a result analysis was conducted in order to compare the results obtained from each of the three categories. They inferred that a combination of features from each of the categories yields better results than features from a single category. In [3], a mix of features were extracted for audio files, first from a set of 10 ragas and then from a set of an additional 10 ragas. They also created separate datasets, one in which the vocals from each of the audio files had been isolated and one in which the vocals were kept. The features from each of these datasets were then used to train three models: ANN, LSTM and XGboost. It was inferred that XGboost provided the best results for both, the dataset containing 10 ragas as well as 20 ragas. However, they also inferred that separating the vocals from the audio files did not yield better results and in fact decreased the accuracies of all the models.

## 4. DATA COLLECTION AND PREPROCESSING

The audio files for this work were procured from the Carnatic Music Raga Recognition Dataset (CMD) of the CompMusic Corpora [8]. This Dataset consists of mp3 audio recordings of performances by various artists in 40 ragas. There are 12 recordings per raga. For our research, we selected 10 of these 40 ragas that are listed in table 1 along with the *swaras*/musical notes that constitute them in both Carnatic and Western notation.

| Raga symbols* | S / C | $R_1$ / $C_\#$ | $R_2$ / D | $G_1$ / $D_\#$ | $G_2$ / E | $M_1$ / F | $M_2$ / $F_\#$ | P / G | $D_1$ / $G_\#$ | $D_2$ / A | $N_1$ / $A_\#$ | $N_2$ / B |
|---|---|---|---|---|---|---|---|---|---|---|---|---|
| At  | ● |   | ● |   | ● | ● |   | ● |   | ● |   | ● |
| Beg | ● |   | ● |   | ● | ● |   | ● |   | ● |   | ● |
| Beh | ● |   | ● |   | ● | ● | ● | ● |   | ● |   | ● |
| Bh  | ● |   | ● | ● |   | ● |   | ● | ● | ● | ● |   |
| Bi  | ● |   | ● |   | ● | ● |   | ● |   | ● |   | ● |
| Dh  | ● | ● |   | ● |   | ● |   | ● | ● | ● |   |   |
| Har | ● |   | ● |   | ● | ● |   | ● |   | ● | ● |   |
| Hu  | ● |   | ● | ● |   | ● |   | ● | ● | ● | ● |   |
| Kal | ● |   | ● |   | ● |   | ● | ● |   | ● |   | ● |
| Kam | ● |   | ● |   | ● | ● |   | ● |   | ● | ● |   |

*Table 1: Raga symbols and the swaras that constitute them, along with their equivalents in western notation*

*Corresponding raga names:*
**At- Atana, Beg- Begada, Beh- Behag, Bh- Bhairavi, Bi- Bilahari, Dh- Dhanyasi, Har- Harikambhoji, Hu- Husseni, Kal- Kalyani, Kam- Kamas**

In order to make the raw audio files suitable for the DL models to be able to comprehend and compute, datasets for our models were created in two ways:

1. **Feature Extraction**: The audio files were reduced to certain features that represent information about their frequencies and magnitudes measured over fixed periods of time. Each audio file was split into 5 second segments. For these segments, the following features were extracted using the Python library Librosa:

- *Mel Frequency Cepstrum Coefficients (MFCC)* : A set of coefficients (usually between 10 and 20) that together constitute a mel-frequency cepstrum, which is a representation of an audio signal in which frequency bands are equally spaced, rather than linearly-spaced. They provide a description of the overall shape of a spectral waveform. For our research, we computed 19 MFCC's as features.

- *Chroma Features:*
  - *Chroma Short-Time Fourier Transform (STFT)*: A 12-element vector representing the energy concentrations of each of the 12 pitch classes in the signal, without taking octaves into consideration.
  - *Chroma Energy Normalised Statistics (CENS)*: A vector that represents statistics of pitch class energy taken over large windows of time in order to smooth local deviations in tempo and articulation. It has proved to be useful for tasks such as audio similarity and matching, which is the theme of a raga recognition problem.

- *Root Mean Square Energy (RMSE):* The square root of mean squared amplitude of the audio signal for an interval of time.

- *Pitch and magnitude*: Two arrays representing the interpolated frequency estimate of a particular harmonic, and the corresponding value of the energy of the peak, respectively.

- *Spectral Centroid:* Center of mass/ frequency band that contains most of the signal's energy

- *Spectral Bandwidth*: Variance from the spectral centroid

- *Rolloff*: The cutoff frequency; the harmonics of frequencies below or above this frequency get filtered out

- *Zero-crossing rate:* The rate at which the signal crosses the zero level axis

Once the features were extracted, they were stored in a Pandas dataframe. The feature columns were assigned as independent variables and the column containing the raga labels was assigned as the dependent variable. The dataset was then split into training, validation and testing sets in the ratio 8:1:1.

2. **Conversion to mel-spectrograms:** A spectrogram is a visual representation of an audio signal that has been subject to a short-time fourier transform. From a spectrogram, one can obtain information of the variation of amplitude of every frequency present in the signal over time. A mel-spectrogram is simply a spectrogram in which the frequencies have been converted to the mel-scale; a scale on which the distances between frequencies are proportional to the human brain's corresponding perception of differences in frequencies.

For the second part of our research, the audio files were converted to mel-spectrograms. Each audio file was again split into 5 second segments which were each converted to mel-spectrograms using the Python library Librosa. They were then stored as image (png) files, which formed the image dataset for our 2-D CNN model. Below is a sample of an image from the dataset.

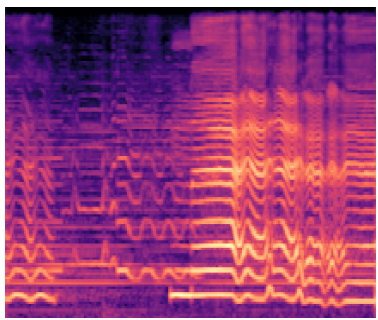

*figure 1: Sample of a mel-spectrogram from an audio clip of the raga Begada*

The images were rescaled to a size of (256,256) and then split into training data, validation data and testing data in the ratio 8:1:1. The images were further rescaled by a factor of 1/.225 and only the training images were augmented. The number of channels for each image was set as 3 and the color mode was set to RGB. [6]

## 5. PROPOSED METHODOLOGY

The proposed methodology is concisely summarized in the form of a flowchart illustrated below:

*figure 2: Flowchart of Proposed Methodology*

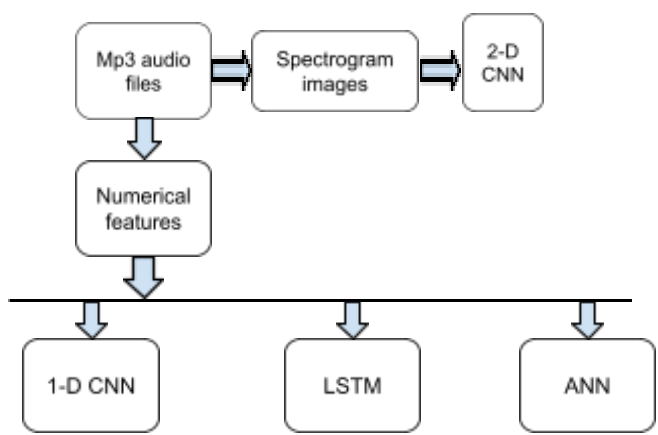

### 5.1 Models used for processing numerical data

*1. 1-D Convolutional Neural Network (1-D CNN)*

The model was designed using one-dimensional CNN layers and dense layers. The input layer is a 1-D CNN layer of size (128,3) which takes an input array of size (1,30). Padding in this layer is set to "same", which ensures that elements are added evenly to all sides of the array when it is being processed by a kernel, thus returning the same output size as its input size. A ReLU activation function is used which is responsible for transforming all the values in the output to the maximum values of either zero or themselves, before it is fed as input to the next layer. Two more such layers were added, of sizes (256,3) and (512,3) respectively. The two dimensional array output from the third CNN layer is then converted to a single linear vector by using a Flatten layer. This vector is then used as input to the consecutive dense layer of size 512. Three more dense layers of sizes 256, 128 and 64 are added respectively. The output layer of the model is a dense layer of size 10, corresponding to the number of classes (ragas) present. It is paired with a softmax activation function, which ensures that the output (prediction) from this layer is an array of size (1,10) in which each element is the probability of its corresponding class. Batch normalization is also used after all the layers. The model is then compiled using the Adam optimizer and sparse-categorical cross entropy loss function. The model is trained with the number of epochs set to 50 and a batch size of 32. The early stopping function is used with patience set to 3. This ensures that if the validation accuracy does not improve for 3 consecutive epochs, training will come to a halt. The training process ended at 11 epochs.

*2. Recurrent Neural Network with Long Short Term-Memory (RNN-LSTM)*

The model consists of two LSTM layers and one dense layer. The input layer is an LSTM layer of size 128 that takes an input array of size (1,30). It has a dropout of 0.05 and a recurrent dropout of 0.25. The consecutive layer is also an LSTM layer of size 64. A Flatten layer is added after the second LSTM layer in order to convert its two dimensional output array to a single linear vector. This is then fed to the output layer of the model, a dense layer of size 10 with a softmax activation function. The model is then compiled using the Adam optimizing function, but this time with a learning rate of 0.0009. The loss function is sparse categorical cross entropy. The model is then trained with the number of epochs set to 100 and a batch size of 32. The early stopping function is used with patience set to 5. The training process ended at 57 epochs.

*3. Artificial Neural Network (ANN)*

The model consists of only dense layers. The input layer has a size of 512 and takes an array of size (1,30) as input. It is paired with the ReLU activation function. Three more such hidden layers are added, with sizes 256, 128 and 64 respectively. The output layer of the model is a dense layer of size 10 paired with the softmax activation function. The model is then compiled using the Adam optimizing function, along with sparse categorical cross entropy as the loss function. The model is trained with the number of epochs set to 50 and a batch size of 32, along with early stopping. The training process ended at 18 epochs.

### 5.2 Model used for processing image data

1. 2-D Convolutional Neural Network (2-D CNN)

The model consists of 2-D CNN layers, Max pooling layers and dense layers. The input layer is a 2-D CNN layer of size (64,3) which takes an input array of size (256,256). Padding in this layer is set to "same", which ensures that pixels are added evenly to all sides of an image when it is being processed by a kernel, thus returning the same output size as its input size. It is paired with a ReLU activation function. Two more such layers were added, of sizes (128,3) and (256,3) respectively. After each of the three 2-D CNN layers, a Batch Normalization layer, a Max Pooling layer of size (2,2) and another Batch Normalization layer are added respectively. The two dimensional array output from the third CNN layer is then converted to a single linear vector by using a Flatten layer. This vector is then used as input to the consecutive dense layer of size 256. Two more dense layers of sizes 128 and 64 are added respectively. The output layer of the model is a dense layer of size 10, paired with a softmax activation function, The model is then compiled using the RMS prop optimizer function and sparse-categorical cross entropy loss function. The model is trained with the number of epochs set to 30 and a batch size of 16, along with early stopping. The training process ended at 12 epochs.

### 6. RESULTS

### 6.1 Results obtained from models that used numerical data

After the models were trained, they were tested with a testing dataset containing 702 samples. In order to calculate the testing accuracy for each model, we calculated confusion matrices as follows:

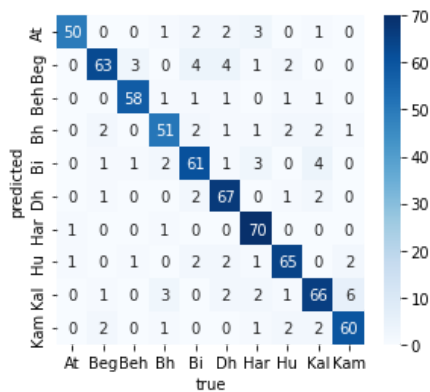

*figure 3:* Confusion Matrix for 1-D CNN model

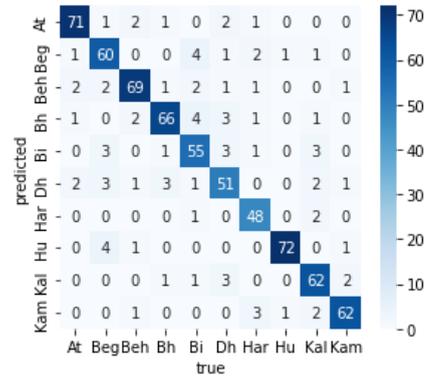

*figure 4:* Confusion Matrix for LSTM model

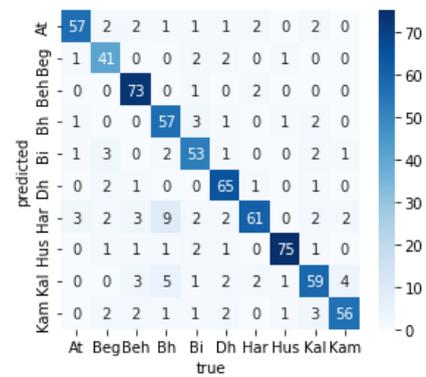

*figure 5:* Confusion Matrix for ANN model

The confusion matrices were plotted with the number of true occurrences of each raga on the x-axis distributed over the number of times each raga was predicted on the y-axis. The elements that constitute the diagonals of each matrix represent the True Positive (*tp*) predictions (of their corresponding ragas) produced by that model. A prediction is said to be *tp* when a class is correctly predicted as present. For a raga on the y-axis, all its corresponding horizontal elements except for that which lies on the diagonal, are False Positive (*fp*) predictions. All cases where a class is incorrectly predicted as present are *fp*. For a raga on the x-axis, all its corresponding vertical elements except for that which lies on the diagonal, are False Negative (*fn*) predictions. All cases where a class is incorrectly predicted as absent are *fn*. For each raga, all elements other than its *tp, fp* and *fn* elements are True Negative (*tn*) predictions. All cases where a class is correctly predicted as absent are *tn*. The overall testing accuracy of each model was calculated as the sum of all correctly predicted values divided by the sum of all predicted values. This is given by:

$$testing\ accuracy = \frac{\Sigma(tp + tn)}{\Sigma(tp + tn + fn + fp)}$$

…..(6.1.1)

Therefore, the testing accuracies in percentage for each model are:

|  | 1-D CNN | LSTM | ANN |
|---|---|---|---|
| Testing accuracy | 97.4% | 97.54 | 97% |

*Table 2: Testing accuracies of models that used numerical data*

In order to analyze the ability of the models to distinguish between 2 different ragas containing the same *swaras*, 2x2 confusion matrices for pairs of these ragas were derived from their corresponding model confusion matrices. The pairs are:

i) Atana and Begada   (ii) Atana and Bilahari

iii) Begada and Bilahari   (iv) Harikambhoji and Kamas

The confusion matrices for these pairs for each model are given below:

|  | 1-D CNN | LSTM | ANN |
|---|---|---|---|
| Pair (i) | At Beg [[50, 0],[0, 63]] At Beg | At Beg [[71, 1],[1, 60]] At Beg | At Beg [[57, 2],[1, 41]] At Beg |
| Pair (ii) | At Bi [[50, 2],[0, 61]] At Bi | At Bi [[71, 0],[0, 55]] At Bi | At Bi [[57, 1],[1, 53]] At Bi |
| Pair (iii) | Beg Bi [[63, 4],[1, 61]] Beg Bi | Beg Bi [[60, 4],[3, 55]] Beg Bi | Beg Bi [[41, 2],[3, 53]] Beg Bi |
| Pair (iv) | Har Kam [[70, 0],[1, 60]] Har Kam | Har Kam [[48, 0],[3, 62]] Har Kam | Har Kam [[61, 2],[0, 56]] Har Kam |

*Table 3: Confusion matrices for pairs of ragas containing the same swaras*

From the confusion matrices above, the misclassification rate was calculated for each pair as the sum of all incorrectly predicted values divided by the sum of all predicted values

$$\text{misclassification rate} = \frac{\Sigma (fn + fp)}{\Sigma (tp + tn + fn + fp)}$$

…(6.1.2)

The misclassification rates for the pairs for each model are tabulated below:

|  | 1-D CNN | LSTM | ANN |
|---|---|---|---|
| Pair (i) | 0% | 1.5% | 2.9% |
| Pair (ii) | 1.76% | 0% | 1.78% |
| Pair (iii) | 3.8% | 5.7% | 4.8% |
| Pair (iv) | 0.7% | 2.65% | 1.68% |

*Table 4: Misclassification rates for the pairs of ragas containing the same swaras*

### 6.2. Results obtained from model that used image data

After the model was trained, it was tested with a testing dataset containing 672 samples. In order to calculate its testing accuracy, a confusion matrix was plotted in the same manner as those plotted for the preceding models

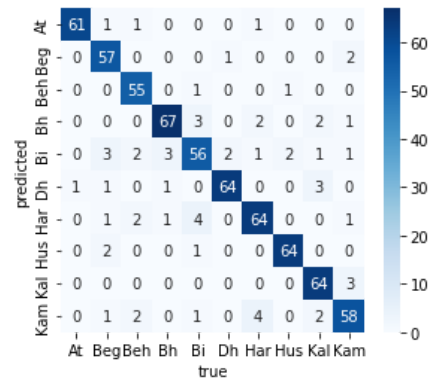

*figure 6: Confusion Matrix for 2-D CNN image classifier mode*

The testing accuracy was calculated using equation 6.1.1 and is given below

|  | 2-D CNN (Image classifier) |
|---|---|
| Testing accuracy | 98.1% |

*Table 5: Testing accuracy for 2-D CNN image classifier model*

In order to analyze the ability of the model to distinguish between 2 different ragas containing the same *swaras*,

2x2 confusion matrices for the same pairs of ragas as in section 6.1 were derived from the model confusion matrix and are tabulated in table 6.. From these confusion matrices, the misclassification rates for these pairs are calculated for the model as in section 6.1 and tabulated in table 7.

|  | 2-D CNN (Image classifier) |
|---|---|
| Pair (i) | At Beg<br>61  1<br>0  57<br>At  Beg |
| Pair (ii) | At Bi<br>61  0<br>0  56<br>At  Bi |
| Pair (iii) | Beg Bi<br>57  0<br>3  56<br>Beg  Bi |
| Pair (iv) | Har Kam<br>64  1<br>4  58<br>Har  Kam |

*Table 6: Confusion matrices of pairs of ragas containing the same swaras for the 2-D CNN model*

|  | 2-D CNN |
|---|---|
| Pair (i) | 0.8% |
| Pair (ii) | 0% |
| Pair (iii) | 2.58% |
| Pair (iv) | 3.9% |

*Table 7: Misclassification rates of the pairs of ragas containing the same swaras for the 2-D CNN model*

### 6.3. Comparison of results

- On comparing accuracies, it was observed that the 2-D CNN **model that used image data yielded greater testing accuracy** than any of the models that used numerical data.

- The misclassification rates for the four pairs of ragas containing the same *swaras* were also compared and it was observed that
  - 
    - For pair i (Atana and Begada) and pair iv (Harikambhoji and Kamas), the 1-D CNN model that used numerical data yielded the lowest misclassification rate.
    - For pair ii (Atana and Bilahari), the 2-D CNN image classifier model and the LSTM model that used numerical data, both yielded the lowest misclassification rate.
    - For pair iii (Begada and Bilahari), the 2-D CNN image classifier yielded the lowest misclassification rate.

### 7. CONCLUSION

The primary objective of this work was to compare two different audio preprocessing techniques; numerical feature extraction, and conversion of audio to images. Three DL models using the first technique and one DL model using the second technique were implemented and their results were analyzed and compared. The model using the second technique yielded state of the art results with a testing accuracy of 98.1%, and proved to exceed the testing accuracies of all three models that used the first technique. **Thus, conversion of audio files to images proved to be a better audio preprocessing technique than extraction of numerical features.**

The secondary objective of this work was to analyze which model could best differentiate between ragas consisting of the same *swaras*. Misclassification rates for four pairs of such ragas were calculated. While the DL models that performed best for each of these pairs were identified, the better preprocessing technique for this task could not be identified as a model using the first technique performed best for two out of the four pairs, while the model using the second technique performed best for the other two pairs (and tied with a model using the first technique for one of these two pairs).

Our research can be explored further by using larger datasets for the same DL models and their corresponding preprocessing techniques in order to check if the second preprocessing technique proves to be better than the first in this case as well. Also, more DL models can be implemented for each preprocessing technique in order to make a more generalized conclusion on which technique gives better results. This might also shed more light on the relationship of misclassification of similar ragas with the DL models and their corresponding preprocessing techniques.